\documentclass[nobm]{epl}
\usepackage{latexsym}
\institute{ 
\inst{1} Department of Microelectronics and Nanoscience, 
Chalmers University of Technology and G\"{o}teborg University, 
S-412 96, G\"{o}teborg, Sweden\\ 
\inst{2} Institute for Physical High Technology, Dept. of Cryoelectronics, 
P.O. Box 100239, D-07702 Jena, Germany\\ 
\inst{3} Institute of Radio Engineering and Electronics, RAS, Moscow 101999,
Russia\\ 
\inst{4} Department of Solid State Physics, Comenius University, 
Mlynsk\'{a} Dolina, SK-842 48 Bratislava, Slovakia\\ 
\inst{5} Department of Applied Physics, Nagoya University, 464-8603, 
Nagoya, Japan\\ 
\inst{6} Electrotechnical Laboratory, Umezono, Tsukuba, Ibaraki, 305-8568, 
Japan 
} 
\input{tcilatex}

\begin{document}

\title{Observation of the second harmonic in superconducting current-phase relation
of Nb/Au/(001)YBa$_{2}$Cu$_{3}$O$_{x}$ heterojunctions}
\author{P. V. Komissinski \thanks{
Corresponding author. E-mail adress: filipp@fy.chalmers.se $<$%
mailto:filipp@fy.chalmers.se$>$}\inst{1,3} \and E. Il'ichev\inst{2} \and G.
A. Ovsyannikov\inst{3} \and S. A. Kovtonyuk \inst{3} \and M. Grajcar\inst{4}
\and R. Hlubina\inst{4} \and Z. Ivanov\inst{1} \and Y. Tanaka \inst{5} \and %
N. Yoshida\inst{5} \and S. Kashiwaya\inst{6}}
\maketitle

\begin{abstract}
The superconducting current-phase relation (CPR) of Nb/Au/(001)YBa$_{2}$Cu$%
_{3}$O$_{x}$ heterojunctions prepared on epitaxial $c$-axis oriented YBa$%
_{2} $Cu$_{3}$O$_{x}$ thin films has been measured in a single-junction
interferometer. For the first time, the second harmonic of the CPR of such
junctions has been observed. The appearance of the second harmonic and the
relative sign of the first and second harmonics of the CPR can be explained
assuming, that the macroscopic pairing symmetry of our YBa$_{2}$Cu$_{3}$O$%
_{x}$ thin films is of the $d+s$ type.
\end{abstract}

\shorttitle{Observation of the second harmonic ...}

\shortauthor{P. V. Komissinski \etal}

\textit{Introduction.} It is now well established that the dominant
component of the superconducting order parameter in the cuprates has $d$%
-wave symmetry (see \cite{Tsuei} and references therein). Moreover, it has
become clear that in orthorhombic materials such as YBa$_{2}$Cu$_{3}$O$_{x}$
(YBCO), a finite component with $s$-wave symmetry is admixed to the dominant 
$d$-wave order parameter. The early in-plane phase sensitive experiments
imply that the $d$-wave component remains coherent through the whole sample 
\cite{Tsuei}, while an elegant $c$-axis tunneling experiment shows directly
that the $s$-wave order parameter component does change sign across the twin
boundary \cite{Kouznetsov}.

The above picture of the YBCO pairing state is challenged by the
experimental observation of a finite $c$-axis Josephson current between
heavily twinned YBCO and a Pb counterelectrode \cite{Sun96}. Namely, the
contribution of the $s$-wave part of the YBCO order parameter to the
Josephson coupling between a conventional superconductor (superconductor
with a pure $s$-wave symmetry of the order parameter, for example Pb or Nb)
and YBCO should average to zero for equal abundances of the two types of
twins in YBCO. In other words, the macroscopic pairing symmetry of twinned
YBCO samples should be a pure $d$-wave \cite{Walker96}. Tanaka has shown
that a finite second order Josephson current obtains for a junction between
the $s$-wave and $c$-axis oriented pure $d$-wave superconductors \cite
{Tanaka94}. However, measurements of microwave induced steps at multiples of 
$hf/2e$ on the $I$-$V$ curves of Pb/Ag/YBCO tunnel junctions imply dominant
first order tunneling \cite{Kleiner}. Therefore the finite $c$-axis
Josephson current has to result from a nonvanishing admixture of the $s$%
-wave component to the macroscopic order parameter of YBCO \cite{Walker96}.
Two alternatives of how this can take place in the junctions based on
twinned YBCO have been discussed in the literature:

(i) Sigrist \textit{et al.} have suggested that the phase of the $s$-wave
component in YBCO does not simply jump from 0 to $\pi $ upon crossing the
twin boundary, but rather changes in a smooth way, attaining the value of $%
\pi /2$ right at the twin boundary \cite{Sigrist96}. The twinned YBCO sample
is thus assumed to exhibit a macroscopic $d+is$ pairing symmetry. A related
picture has been proposed by Haslinger and Joynt, who suggest a $d+is$
surface state of YBCO \cite{Haslinger}.

(ii) A difference in the abundances of the two types of twins implies a $d+s$
symmetry of the macroscopic pairing state \cite{ODonovan97}. Let us point
out that also structural peculiarities of other type (such as a lamellar
structure in a preferred direction) may lead to the $d+s$ macroscopic
pairing symmetry.

In this paper we report the observation of a large second harmonic of the
Josephson current in Nb/Au/(001)YBCO junctions. By comparing the relative
signs of the first and second harmonics of the Josephson current, we show
that the macroscopic $d+s$ pairing symmetry is realized in our YBCO samples.
Our results might be relevant also in search for so called ''quiet'' qubits
which can be realized making use of junctions with a dominant second
harmonic of the Josephson current \cite{Ioffe99}. Another promising route to
fabricate such junctions is via 45$^{\circ }$ grain boundary Josephson
junctions \cite{Ilichev99,Ilichev01a}.

\textit{Experimental.} In realization of the Josephson junctions between
conventional superconductors and $c$-axis oriented high-temperature ones an
extensive use has been made of the Pb/Ag/(001)YBCO heterojunctions \cite
{Sun96}. In search for a combination of the superconducting counterelectrode
and a normal-metal buffer layer with the highest possible interface quality,
in the present work we have decided to study the Nb/Au/(001)YBCO
heterojunction. Our epitaxial (001)-oriented YBCO films with thicknesses of
150~nm were obtained by laser deposition on (100) LaAlO$_{3}$ and (100) SrTiO%
$_{3}$ substrates. The films are usually twinned in the $ab$-plane. The
superconducting transition temperature of our YBCO films was determined by
magnetic susceptibility measurements as $T_{c}=88\div 90$ K. The YBCO films
were \textit{in situ} covered by a $8\div 20$ nm thick Au layer, thus
preventing the degradation of the YBCO surface during processing.
Afterwards, $200$~nm thick Nb counterelectrodes were deposited by
DC-magnetron sputtering. Photolithography and low energy ion milling
techniques were used to fabricate the Nb/Au/YBCO junctions. The interface
resistance per unit area $R_{B}=R_{N}S$ (where $R_{N}$ is the normal state
resistance and $S$ is the junction area) was $R_{B}=10^{-5}\div 10^{-6}\
\Omega \cdot $cm$^{2}$. Details of the junction fabrication were reported
elsewhere \cite{KomissinskiJETP}.

Surface quality of the YBCO films is very important when current transport
in the $c$-axis direction is investigated. High-resolution atomic force
microscopy reveals a smooth surface consisting of approximately 100~nm long
islands with vertical peak-to-valley distance of $\approx{3}$~nm (fig.~1).
We can exclude that substantial $ab$-plane tunnel currents flow between YBCO
and Nb at the boundaries of these islands. In fact, theory predicts
formation of midgap states at the surface of semi-infinite CuO$_{2}$ planes 
\cite{Hu,Tanaka95}. Therefore zero bias conductance peaks should be expected
in the $I$-$V$ characteristics at temperatures larger than the critical
temperature of Nb, if the contribution of $ab$-plane tunneling was
nonnegligible. However, no such peaks have been observed for all fabricated
Nb/Au/YBCO junctions. Moreover, from the size of the islands and from the
vertical peak-to-valley distance we estimate that the area across which $ab$%
-plane tunneling might take place is less than 6\% of the total junction
area. Since the interface resistances per square are of the same order of
magnitude \cite{Sun96} for both, the $c$-axis and $ab$-plane junctions, we
conclude that $ab$-plane tunneling from YBCO, if present, is negligibly
small.

We have measured more than 20 single junctions with areas in the range from $%
10\times 10\ \mu $m$^{2}$ to $100\times 100\ \mu $m$^{2}$. At small voltages
the typical $I$-$V$ curves can be described by the resistively shunted
junction model with a small capacitance \cite{Likharev}. Typical critical
current densities were $j_{c}=1\div 12$~A/cm$^{2}$ and $j_{c}R_{B}=10\div
90\ \mu $V. The differential resistance $vs.$ voltage dependence $R_{d}(V)$
exhibits a gap-like structure at $V\approx 1.2$~mV at $T=4.2$ K (see
fig.~2). This structure has a BCS-like temperature dependence and disappears
at $T_{cR}\approx 9.1$~K, therefore we ascribe it to the superconducting
energy gap of Nb.

The CPR measurements were performed using a single-junction interferometer
configuration in which a junction of interest is inserted into a
superconducting loop with an inductance $L\approx 80$~pH. We measure the
impedance of a parallel resonance circuit inductively coupled to the
interferometer as a function of the external magnetic flux $\Phi_{e}$
threading the interferometer. The dimensionless CPR $f(\varphi
)=I_{s}(\varphi )/I_{c}$ (where $I_{s}(\varphi)$ is the Josephson current)
can be extracted from the following equations: 
\begin{eqnarray}
\varphi&=&\varphi_{e}-\beta f(\varphi),  \label{eq:phi} \\
\tan\alpha&=&\frac{k^{2}Q\beta f^{\prime }(\varphi )} {1+\beta f^{\prime
}(\varphi )},
\end{eqnarray}
where $\varphi_{e}=2\pi\Phi_{e}/\Phi_{0}$, $\varphi$ is the phase difference
across the Josephson junction, the prime denotes a derivative with respect
to $\varphi$, $\alpha$ is the phase shift between the driving current and
the tank voltage at the resonant frequency, $\beta =2\pi LI_{c}/\Phi_{0}$ is
a normalized critical current, $Q$ is the quality factor of the parallel
resonance circuit, $k$ is the coupling coefficient between the RF SQUID and
the tank coil, and $\Phi_{0}$ is the flux quantum. This method, being
differential with respect to $\varphi$, provides a high sensitivity of the
CPR measurement. Moreover, $I_s(\varphi)$ is measurable even if the thermal
energy exceeds the Josephson coupling energy. In fact, critical currents
down to 50 nA were recently detected at $T=4.2$ K \cite{Ilichev01}.

For the junction SQ10, $\beta $ varied between 0.4 and 0.27 for temperatures
in the range $T=1.7\div 6.0$~K. Since $\beta <1$, we can extract the CPR
from the $\alpha (\varphi _{e})$ dependence for the complete phase range
(see fig. 3). Fourier analysis of the experimentally obtained CPR shows
substantial first and second harmonics and negligibly small higher-order
harmonics. Therefore we can write 
\begin{equation}
I_{s}(\varphi )=I_{1}\sin \varphi +I_{2}\sin 2\varphi  \label{eq:Fourier}
\end{equation}
for all temperatures below the transition temperature of Nb. The sign of $%
I_{2}$ is always opposite to that of $I_{1}$. In what follows we use the
convention that $I_{1}>0$ and therefore $I_{2}<0$. The ratio $|I_{2}/I_{1}|$
grows with decreasing temperature reaching $|I_{2}/I_{1}|\approx 0.16$ at $%
T\approx 1.7$~K, when $I_{1}=1.57$ $\mu $A and $I_{2}=-0.25$ $\mu $A (see
inset to a fig. 3(b)).

We point out that the main result of this paper, namely the large negative
value $I_{2}/I_{1}\approx -0.16$ observed at low temperatures, is not the
result of an indirect data analysis, but it follows directly from the
measured $\alpha (\varphi _{e})$ dependences. In fact, one finds readily
that 
\begin{equation}
{\frac{d^{2}\tan \alpha }{d\varphi _{e}^{2}}}=\frac{k^{2}Q\beta f^{\prime
\prime \prime }}{(1+\beta f^{\prime })^{4}},
\end{equation}
where all derivatives are taken at $\varphi =0$ or $\varphi _{e}=0$. Thus,
the existence of the local minima of $\alpha (\varphi _{e})$ at multiples of 
$\pi $ dictates that $f^{\prime \prime \prime }>0$, or, making use of Eq.~(%
\ref{eq:Fourier}), $I_{2}/I_{1}<-1/8$. Note that neither the conventional
tunneling theory nor the I and II theories by Kulik and Omelyanchuk predict
such local minima on the derivatives of CPR \cite{Likharev}.

\textit{Discussion.} Let us estimate the transparency of the barrier between
YBCO and Nb from the normal-state resistance per unit area $R_B$. According
to the band-structure calculations (for a review, see \cite{Pickett89}), the
hole Fermi surface of YBCO is a slightly warped barrel with an approximately
circular in-plane cross-section (to be called Fermi line) with radius $k_F$.
In what follows, we represent the electron wavevector $\mathbf{k}$ in
cylindrical coordinates, $\mathbf{k}=(k,\theta,k_z)$. We estimate the
uncertainty of the in-plane momentum as $\delta k\approx 2\pi/l$, where $l$
is the characteristic size of the islands on the YBCO surface (fig.1). We
evaluate $R_B$ making use of the Landauer formula and note that only
tunneling from a shell around the Fermi line with width $\delta k$ is
kinematically allowed. The barrier transparency $D(\theta)$ depends on the
details of the $c$-axis charge dynamics in YBCO, with maxima in those
directions $\theta$, in which the YBCO $c$-axis Fermi velocity $w(\theta)$
is maximal. Since for $\theta=\pi/4$ and symmetry equivalent directions $%
w(\theta)$ is minimal \cite{Xiang96}, we expect that there will be 8 maxima
of $D(\theta)$ on the YBCO Fermi line where $D(\theta)\approx D$, which are
situated at $\theta=\theta_0$ and symmetry equivalent directions. The
modulation of the function $D(\theta)$ depends on the thickness of the
barrier between YBCO and Nb \cite{Wolf85}. We consider two limiting
distributions of the barrier transparency $D(\theta)$ along the YBCO Fermi
line: (a) a featureless $D(\theta)\approx D$ and (b) a strongly peaked $%
D(\theta)$, roughly corresponding to thin and thick barriers, respectively 
\cite{Wolf85}. In the thick barrier limit the angular size of the maxima of $%
D(\theta)$ can be estimated as $\delta\theta\approx \delta k/k_F$. With
these assumptions we find 
\begin{equation}
R_B^{-1}={\frac{\langle D\rangle e}{\Phi_0}}A,  \label{eq:conductivity}
\end{equation}
where $A$ measures the number of conduction channels and $%
\langle\ldots\rangle$ denotes an average over the junction area. In the thin
and thick barrier limits, we find $A\approx k_F\delta k/\pi$ and $A\approx
2\delta k^2/\pi$, respectively. Taking $l\approx 100$ nm and $k_F\approx 0.6$
\AA$^{-1}$ \cite{Shen95}, the measured $R_B=6\times 10^{-5}$ $\Omega$cm$^2$
of the junction SQ10 can be fitted with $\langle D\rangle_{\mathrm{thin}
}\approx 1.7\times 10^{-5}$ and $\langle D\rangle_{\mathrm{thick}}\approx
8.3\times 10^{-4}$.

Since we have observed no midgap surface states in the $R_d(V)$ curves, we
can neglect the surface roughness, and the Josephson current can be
calculated from \cite{Zaitsev84} 
\begin{equation}
I_s(\varphi)={\frac{2e}{\hbar}}\sum_{k,\theta} k_BT\sum_{\omega} {\frac{
D\Delta_R\Delta_{\mathbf{k}}\sin\varphi}{2\Omega_R\Omega_{\mathbf{k}}+D\left[
\omega^2+\Omega_R\Omega_{\mathbf{k}} +\Delta_R\Delta_{\mathbf{k}}\cos\varphi %
\right]}},
\end{equation}
where the sum over $k,\theta$ is taken over the same regions with areas $A$
as in Eq.~(\ref{eq:conductivity}), $\Delta_R$ and $\Delta_{\mathbf{k}}$ are
the Nb and YBCO gaps, respectively, and $\Omega_i=\sqrt{\omega^2+\Delta_i^2}$
with $i=R,\mathbf{k}$. Keeping only terms up to second order in the (small)
junction transparency $D$, the Josephson current densities $j_i=I_i/S$ read 
\begin{eqnarray}
j_1(T)R_B&\approx&{\frac{\Delta_s}{\Delta_d^\ast}}{\frac{\Delta_R(T)}{e}},
\label{eq:j_1} \\
j_2(T)R_B&\approx&-{\frac{\pi}{8}} {\frac{\langle D^2\rangle}{\langle
D\rangle}} {\frac{\Delta_R(T)}{e}}\tanh\left({\frac{\Delta_R(T)}{2k_BT}}
\right),  \label{eq:j_2}
\end{eqnarray}
where $\Delta_d^\ast=\pi\Delta_d[2\ln(3.56\Delta_d/T_{cR})]^{-1}$ and $%
\Delta_d^\ast=\Delta_d|\cos 2\theta_0|$ in the thin and thick barrier
limits, respectively. In Eqs.~(\ref{eq:j_1},\ref{eq:j_2}) we used the YBCO
gap $\Delta(\theta)=\Delta_s+\Delta_d\cos 2\theta$, where $\Delta_d$ and $%
\Delta_s$ are the $d$-wave and $s$-wave gaps. We have assumed that $%
\Delta_d^\ast$ is larger than both, $\Delta_R$ and $\Delta_s$. The factor $%
\Delta_s/\Delta_d^\ast$ can be estimated from the measured $j_1R_B$ products
for Josephson junctions between untwinned YBCO single crystals and Pb
counterelectrodes. For such junctions $j_1(0)R_B\approx 0.5\div 1.6$~meV 
\cite{Sun96}. Using the Pb gap $\Delta_R=1.4$~meV in Eq.~(\ref{eq:j_1}), we
obtain $\Delta_s/\Delta_d^\ast\approx 0.36\div 1.1$.

Note that within the above microscopic $d+s$ scenario, the signs of $I_{1}$
and $I_{2}$ are different. This feature remains valid also in the
macroscopic $d+s$ scenario, whereas within the macroscopic $d+is$ picture,
the same signs of $I_{1}$ and $I_{2}$ are expected. Thus we conclude that
the finite first harmonic has to be due to the macroscopic $d+s$ symmetry of
our YBCO sample. In fact, detailed structural studies show that for
sufficiently thin YBCO films, the abundances of the two types of twins can
be different even for films grown on the cubic substrate SrTiO$_{3}$\cite
{Didier97}. If we denote the twin fractions as $(1+\delta )/2$ and $%
(1-\delta )/2$, then the measured first harmonic of the CPR, $\langle
j_{1}\rangle $, is proportional to the deviation from equal population of
twins, $\langle j_{1}\rangle =\delta j_{1}$ \cite{ODonovan97}. Using $\Delta
_{R}=1.2$~meV determined from the $R_{d}(V)$ data and our estimate $\Delta
_{s}/\Delta _{d}^{\ast }\approx 0.36\div 1.1$, we find that the measured
first harmonic $\langle j_{1}\rangle $ for the junction SQ10 can be fitted
with $\delta \approx 0.07\div 0.21$, which is in qualitative agreement with 
\cite{Didier97}, where $\delta \approx 0.14$ for 1000 \AA\ thick YBCO films
has been observed.

Fitting the measured $j_2R_B$ of the junction SQ10 by Eq.~(\ref{eq:j_2}) we
obtain $\langle D^2\rangle/\langle D\rangle\approx 3.2\times 10^{-2}$, which
is much larger than both $\langle D\rangle_{\mathrm{thin}}$ and $\langle
D\rangle_{\mathrm{thick}}$. This diference can be explained provided the
junction transparency $D$ is a fluctuating function of the position $\mathbf{%
r}$. In fact, adopting the WKB description of tunneling \cite{Wolf85}, we
write $D(s(\mathbf{r}))=\exp(-s_0-s(\mathbf{r}))$, where $s_0$ is the WKB
tunneling exponent and $s(\mathbf{r})$ its local deviation from the mean.
Assuming a Gaussian distribution of $s$ with a mean deviation $\eta$, $%
P(s)\propto\exp(-s^2/\eta^2)$, we estimate the spatial averages as $\langle
D^n\rangle=\int_{-s_0}^{s_0} ds P(s) D^n(s)$. In the thin barrier limit, the
values $\langle D^2\rangle_{\mathrm{thin}}=8.6\times 10^{-7}$ and $\langle
D\rangle_{\mathrm{thin}}=2.3\times 10^{-5}$ required to fit the experiments
correspond to an average WKB exponent $s_0^{\mathrm{thin}}\approx 15.5$ with 
$\eta_{\mathrm{thin}}\approx 4.3$. In the thick barrier limit we obtain $%
s_0^{\mathrm{thick}}\approx 9.1$ and $\eta_{\mathrm{thick}}\approx 2.8$.

Let us consider also the second harmonic generation by a mechanism proposed
by Millis for planar junctions \cite{Millis94}. We can view the junction as
a checkerboard of 0 and $\pi$ junctions with a lattice constant $a$
(characteristic size of an YBCO twin) and local critical current density $%
j_1 $. As shown by Millis, spontaneous currents are generated in the ground
state of such a junction and the junction energy is minimized for the phase
difference $\pm\pi/2$. An explicit calculation in the limit $%
a,\lambda_R\ll\lambda\ll\lambda_c$ (where $\lambda,\lambda_c$ are the $ab$%
-plane and $c$-axis penetration depths of YBCO and $\lambda_R$ is the Nb
penetration depth) yields 
\[
j_{2,\mathrm{Millis}}\approx -{\frac{1}{4\sqrt{2}}} {\frac{
\mu_0j_1^2a\lambda\lambda_c}{\Phi_0}}. 
\]
Comparing $j_{2,\mathrm{Millis}}$ with Eq.~(\ref{eq:j_2}), we find 
\[
{\frac{j_{2,\mathrm{Millis}}}{j_2}}\approx {\frac{8\sqrt{2} F}{\pi^2}}\left({%
\frac{\Delta_s}{\Delta_d^\ast}}\right)^2 {\frac{a}{l}}{\frac{\Delta_R}{\hbar
c/\lambda_c}} {\frac{E_0}{\hbar c/\lambda}}, 
\]
where $E_0=e^2q_0/4\pi\epsilon_0$ with $q_0=k_F$ and $2\delta k$ for a thin
and thick barrier, respectively. $F$ depends on the characteristic length
scale $L$ of the fluctuations of the junction transparency: for $L\ll a$, $%
F\approx \langle D\rangle^2/\langle D^2\rangle$, whereas for $L\gg a$, $%
F\approx 1$. In order to estimate the upper bound of the ratio $j_{2,\mathrm{%
Millis}}/j_2$, we consider the thin barrier limit, take $F\approx 1$, $%
\Delta_s/\Delta_d^\ast\approx 1.1$, $a\approx 10$~nm, $\lambda\approx 240$%
~nm, and $\lambda_c\approx 3$~$\mu$m (the last two values are valid for
underdoped YBCO \cite{Cooper94}) and we find that $j_{2,\mathrm{Millis}%
}/j_2<0.03$. Thus we can safely neglect the contribution of the Millis
mechanism to the second harmonic of the CPR.

\textit{Conclusion.} We have observed the second harmonic $I_{2}$ of the
current-phase relation in $c$-axis Nb/Au/YBCO heterojunctions. We have shown
that the relative phases of the first and second harmonics of the CPR
together with their amplitudes $I_{1}$ and $I_{2}$ and the normal-state
resistance of the Josephson junctions observed in the experiment can be
explained making use of a single set of parameters within the standard
microscopic $d+s$ picture of the YBCO pairing state, assuming the difference
in the abudances of the two types of twins and that the barrier fluctuates
along the junction.

\acknowledgments We would like to thank P. Dmitriev for the Nb films
deposition, P. Mozhaev and K. I. Constantinyan for assistance in
measurements, M. Yu. Kupriyanov, T. L\"{o}fwander, V. Shumeiko, A.
Tzalenchuk, A. V. Zaitsev for fruitful discussions, and prof. T. Claeson for
a critical reading of the manuscript. This work is supported by the INTAS
program of the EU (Grant No.~11459), the DFG (Ho461/3-1), the Swedish
Material Consortium of superconductivity, the Russian Foundation of
Fundamental Research, and the Russian National Program on Modern Problems of
Condensed Matter. Partial support by the $D$-wave Systems, Inc., by the NATO
Science for Peace Program N97-3559, and by the Slovak Grant Agency VEGA
(Grant No.~1/6178/99) is gratefully acknowledged.

Figure captions

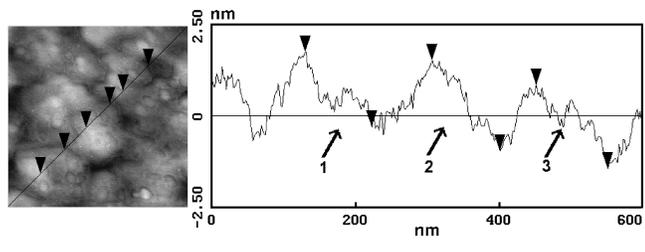
Fig. 1. Left panel: A high resolution AFM image of the surface of a 150 nm
thick YBCO film. Right panel: Height profile of the YBCO surface along the
line indicated in the left panel. The peaks and valleys are shown by
markers. The peak-to-valley distance is $\approx $3 nm and about 100 nm in
the vertical and horisontal directions respectively.

Fig.2(a) Typical $I-V$ curve and (b) differential resistance vs. voltage
dependence $R_{d}(V)$ of a Nb/Au/YBCO junction with a 20~nm thick Au film,
measured at $T=4.2$~K. The low voltage part of the $I-V$ curve ( $V\leq 0.3$%
~mV) is shown on the inset.

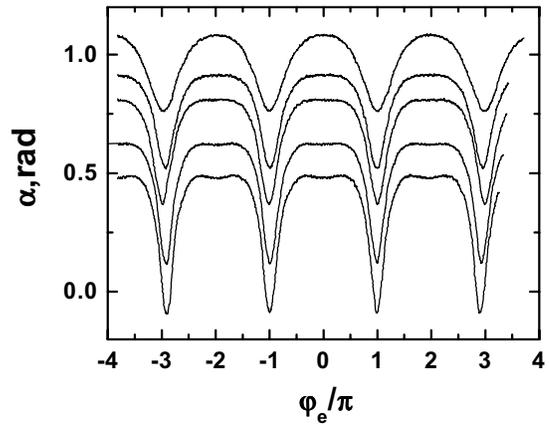
Fig. 3. Phase shift $\alpha $ as a function of $\varphi _{e}$ for the
junction SQ10 at $T=$1.7, 2.5, 3.5, 4.2, and 6.0~K (from bottom to top).

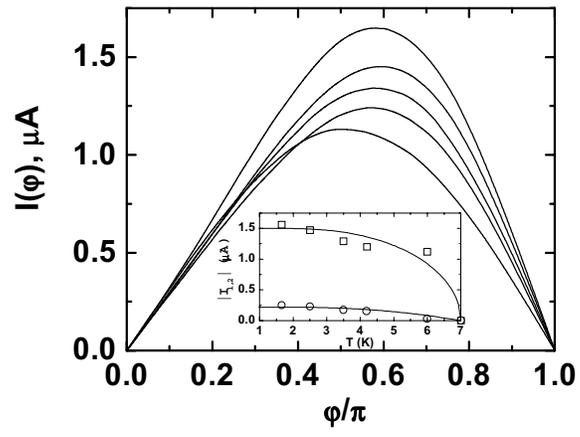
Fig. 4. The current-phase relation $I(\varphi )$ of the junction SQ10 at $T$%
=1.7, 2.5, 3.5, 4.2, and 6.0~K (from top to bottom). Inset: Temperature
dependence of $I_{1}$ (squares) and $|I_{2}|$ (circles). Solid lines are
fits to Eqs.~(\ref{eq:j_1},\ref{eq:j_2}) using $\Delta _{R}(T)=\Delta
_{R}(0)\tanh [\Delta _{R}(T)T_{cR}/\Delta _{R}(0)T]$.

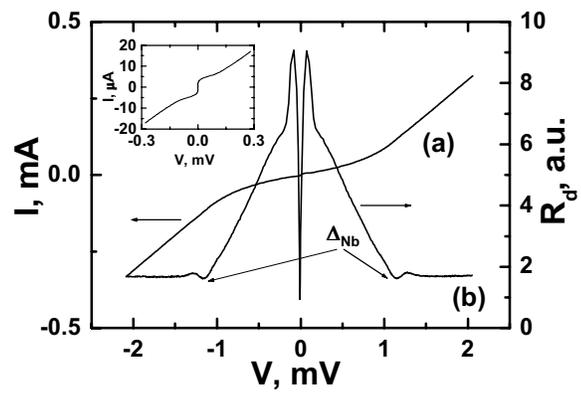
\end{document}